\documentclass[10pt]{article}
\usepackage{sao2}
\usepackage{psfig,epsf}

\issue{2011, 66, 199--206}

\def\apj{Astrophys. J}

\def\apjs{Astrophys. J. Supp.}

\begin{document}
\markboth{O.V.~Verkhodanov,M.L.~Khabibullina }
{STATISTICS OF WMAP ILC MAP TEMPERATURE FLUCTUATIONS}
\title{Statistics of WMAP ILC Map Temperature Fluctuations
Towards Distant Radio Galaxies}

\author{O.V.~Verkhodanov\inst{a},
M.L.~Khabibullina\inst{a}
}
\institute{
$^a$\saoname}

\date{September 6, 2010}{February 1, 2011}
\maketitle
\begin{abstract}
For 2442 galaxies of the catalog, compiled based on the NED, SDSS,
and CATS survey data with redshifts $z>0.3$ we conducted an
analysis of the amplitude of temperature fluctuations in the
cosmic microwave background (CMB) in the points, corresponding to
the direction to these objects. To this end, we used the ILC map
from the WMAP mission seven-year data release. We have estimated
the dipole component of the background and tested the hypothesis
of Kashlinsky on the existence of a ``dark bulk flow'', determined
for the estimated dipole component of the CMB WMAP by the value of
the CMB anisotropy in the direction to the clusters of galaxies.
We show that the amplitude of this dipole $T_{max}=0.012$\,mK is
located within the $\sigma$ interval, estimated by Monte Carlo
simulations for the Gaussian fluctuations of the CMB signal in the
$\Lambda$CDM model. The low amplitude of the dipole indicates that
it is impossible to confirm this hypothesis from the WMAP data. In
addition, we studied the statistics of the fluctuation amplitude
of the microwave signal in the direction to radio galaxies. A
weakening of the absolute value of the mean signal in the
corresponding fields was discovered.
\end{abstract}

\maketitle

\section{INTRODUCTION}

The studies of properties of the large-scale structure, reflected
in the statistics of the  Cosmic Microwave Background (CMB), are
of particular interest in connection with the advent of new
high-accuracy  observations in the millimeter and submillimeter
ranges
\cite{wmapresults:verkh2,wmap3ytem:verkh2,wmap5ytem:verkh2,wmap7ytem:verkh2}.
Using the data on galaxy clusters and the magnitude of the CMB
signal fluctuations in the direction to their centers, where the
signal fluctuations are caused by the Sunyaev-Zel'dovich  effect
(SZ) \cite{zs:verkh2}, we can estimate the cosmological parameters
(the Hubble constant $H_0$, and the density parameter $\Omega_0$)
\cite{dezotti:verkh2}. In addition, one of the components of the
SZ effect---the kinematic SZ effect, caused by the proper motion
of the cluster as a whole with respect to the CMB reference
system---opens the possibility of studying the large-scale fluxes
of matter. In particular, \cite{kash1:verkh2,kash2:verkh2}
describe the study of large-scale ``dark bulk flows'' in motion of
clusters of galaxies, based on the WMAP5~ data and the catalog,
compiled from the X-ray data and containing more than 1000
clusters. The authors have calculated the dipole from the CMB
pixel values. The dipole is determined, in their opinion, by the
kinematic SZ effect; it reveals a constant velocity, at least on
the scale of 800\,Mpc.

Note that despite the low WMAP sensitivity to the SZ effect due to
the relatively poor resolution of the final map of the isolated
CMB signal (the limiting spherical harmonics---a multipole---in
the first WMAP releases for the ILC CMB map there is
$\ell_{max}\sim100$, or, in terms of resolution,
$\theta\sim$55\arcmin\, and in the seven-year release it is
$\ell_{max}\sim150$ ($\theta\sim$35\arcmin), while the SZ effect
for most of galaxy clusters is strongly manifested on the small
scale: $\theta<$10\arcmin), the discussed effect is of great
interest in terms of its manifestation in the ongoing Planck
mission\footnote{\tt http://www.rssd.esa.int/Planck/}.

Also note that despite the limitations on resolution (channel W
has a better resolution of $12.6'$ with the frequency of 94 GHz)
and on the sensitivity of WMAP maps, various attempts to measure
the SZ effect statistically have been made both in the population
averaging (``stacking'') (see, e.g., recent papers
\cite{diego_partridge:verkh2,wmap7cosmo:verkh2}),
 and directly for individual bright clusters, such as Coma~\cite{wmap7cosmo:verkh2},
where the effect was observed. Diego and
Partridge~\cite{diego_partridge:verkh2} used  three  channels: Q
(43 GHz, ${\rm FWHM} = 0.51^\circ$), V (61  GHz, \linebreak
0.35$^\circ$) and W (94	 GHz, 0.22$^\circ${}) and obtained
averaged profiles in the direction to known clusters. In
particular, besides the presence of the desired signal and the
growth of its amplitude with  frequency, the authors found that
the signal's value is lower than expected from the X-ray data, the
fact that they linked with the presence of a point source in the
cluster. The WMAP team~\cite{wmap7cosmo:verkh2} has conducted a
similar study, using two channels: V and W. They also applied the
averaging of different regions in the direction to known clusters
with X-ray emission and evidenced the effect. In addition, they
discovered that the effect can be traced down to the scales of
$\theta=1.05^\circ$. The signal on the scale of
$\theta=0.58^\circ$ is considered as true, and that on the scale
of $\theta=1.05^\circ$---as a statistical fluctuation. At that,
the authors \cite{wmap7cosmo:verkh2} have concluded that the
apparent effect for the averaged sources is consistent with that,
expected from the X-ray observations. Let us emphasize that the
evaluations are made for the source, producing the SZ effect, and
averaged over the sky. Recent experiments, such as the the one
conducted on the Atacama Cosmology Telescope
(ACT)~\cite{act_sz:verkh2} and Planck
mission~\cite{planck_sz:verkh2} already allow to isolate and see
this effect directly.

Despite the fact that for most clusters the SZ effect is not
directly visible in the WMAP data, it is estimated statistically.
We believe that the approach proposed in
\cite{kash1:verkh2,kash2:verkh2} deserves special attention. We
develop it in this paper, using the amplitude of temperature
fluctuations in the points, corresponding to the direction to
distant radio galaxies from our survey for the analysis. To detect
the possible ``dark bulk flow'' we need to evaluate the signal
towards distant radio galaxies. A useful moment is the fact that
the CMB signal in the ILC map of fluctuations is isolated using
the channels, in which the SZ effect, if existing, will have a
negative signal. Fortunately, the angular scale of the SZ effect,
discovered by the WMAP \mbox{team \cite{wmap7cosmo:verkh2}}
corresponds to the resolution of the ILC map  of seven-year
observations.

In the standard scheme of galaxy formation, the radio source
lights up as a result of a merger of galaxies, and formation of an
accretion disk and jets, observed at radio and other wavelengths.
As a rule (see reviews in
\cite{vo_par3:verkh2,miley_debreuck:verkh2}), the most powerful
radio galaxies, visible at large redshifts are identified with
giant elliptical galaxies, which are mainly the central galaxies
of clusters and are formed by merging. This property (radio
emission) can be used to find distant clusters and protoclusters
of galaxies. For example, \cite{venemans:verkh2} presents the
results of a program, conducted with the ESO VLT telescope,
searching for the emerging clusters of galaxies near the powerful
radio galaxies at redshifts $2<z<5.2$ with the radio fluxes
$B_{2.7\,\rm GHz}>10^{33}$\,erg s$^{-1}$ Hz$^{-1}$ sr$^{-1}$. The
authors have selected 150 objects and examined the fields around
nine of them. In the fields the authors have selected the
galaxies, emitting in  Ly$\alpha$ (the so-called
Ly$\alpha$-emitters), the redshifts of which were measured. Using
the data on spatial density of objects it was concluded whether
they belong to protoclusters. The size of such protoclusters is at
least 1.75\,Mpc. It was shown that 75\% of radio galaxies with
$z>2$ are associated with protoclusters. Hence, we can estimate
that approximately 3$\times$10$^{-8}$  of the forming clusters lay
in the interval of $2<z<5.2$ in the comoving unit volume with a
side of 1\,Mpc with an active radio source. But it is likely that
within a given range of redshifts the number of protoclusters may
be higher, since an active radio source may simply not be
observed.

Using the survey of remote (for example, \mbox{$z>0.3$}) radio
galaxies, we can track the locations of clusters and protoclusters
of galaxies, which in turn may affect the fluctuations of the
microwave background. A sample of radio galaxies, prepared by us
\mbox{earlier
\cite{rg_list1:verkh2,rg_list2:verkh2,rg_list3:verkh2}} is very
suitable to this end, and we use it in our study.

\section{DATA ANALYSIS}
\subsection{Input Arrays}

A sample of radio galaxies with $z>0.3$
\cite{rg_list1:verkh2,rg_list2:verkh2,rg_list3:verkh2,rg_list4:verkh2,rg_list5:verkh2}
\linebreak  was constructed using the tools and archives of the
following databases: NED\footnote{\tt
http://nedwww.ipac.caltech.edu}, CATS\footnote{\tt
http://cats.sao.ru} \cite{cats1:verkh2,cats2:verkh2}, \linebreak
SDSS\footnote{\tt http://www.sdss.org} \cite{sdss:verkh2} with the
aim to use it in various statistical and cosmological tests
\cite{vo_par3:verkh2,vo_par1:verkh2,vo_par2:verkh2},  which
require an  analysis of a large sample of objects of the same
nature. The NED database was used to construct the primary list of
objects. We selected from it the objects with required parameters,
mainly based on redshift ($z>0.3$) and morphological properties of
radio galaxies. The initial catalog contained 3364 objects. Such a
sample of galaxies is contaminated with objects with incomplete
information, or objects with other properties. Therefore, special
attention was paid to clean the original sample from the odd
sources: galaxies (1) with redshifts, determined by the
photometric method; (2) those with quasar properties from the
available literature data. The final catalog contains 2442 sources
with spectroscopic redshifts, the photometric values and flux
densities in the radio range, the sizes of radio sources, as well
as radio spectral indices, which were calculated based on the
results of cross-identification with the radio catalogs, stored in
the CATS, in the frequency range from 325\,MHz to 30\,GHz. The
positions of 2442 galaxies on the sphere are demonstrated in
Fig.\,1.

\begin{figure*}[!th]
\centerline{\hbox{
\psfig{figure=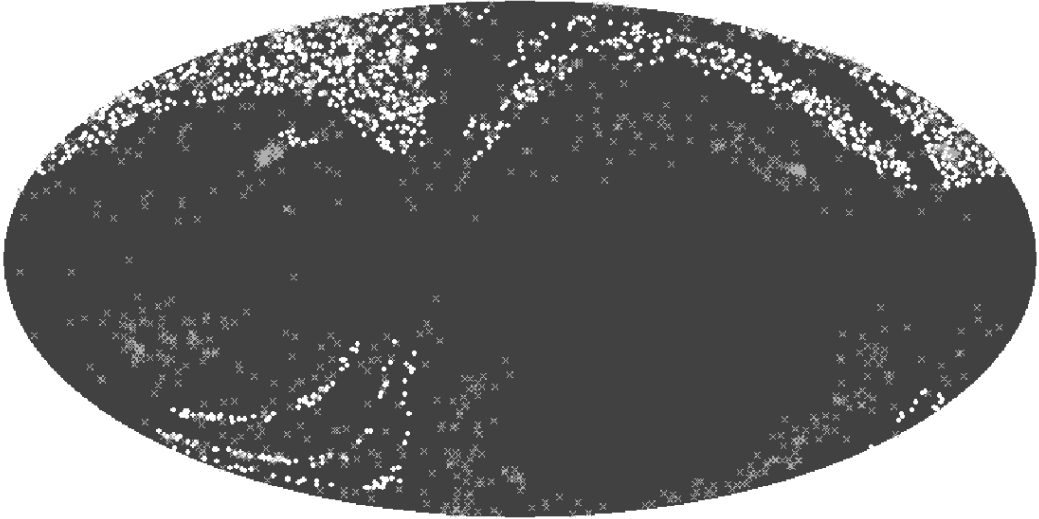,width=14cm}
}} \caption{Position of selected radio sources on the celestial
sphere in galactic coordinates. White circles mark the SDSS
objects, gray crosses---the remaining sources.} \label{f1:verkh2}
\end{figure*}

As the CMB map, we used the map of the distribution of the cosmic
microwave background anisotropy, reconstructed based on the
multi-frequency observations via the technique of internal linear
combination (ILC) of the  background
components~\cite{wmapresults:verkh2} in the observational program
of the WMAP space mission\footnote{\tt
http://lambda.gsfc.nasa.gov} (Wilkinson Microwave Anisotropy
Probe). To construct the WMAP map, the data in five bands were
used: 23\,GHz (K-band), 33\,GHz (Ka-band), 41\,GHz (Q-band),
61\,GHz (V-band), and 94\,GHz (W-band). The ILC map contains
information on the distribution of the microwave background for
not very high harmonics ($\ell\le150$).

Nevertheless, it is currently the most tested and used CMB map.
For the data analysis, we use the distribution of the signal
$\Delta T(\theta,\phi)$,  describing the temperature anisotropy on
the sphere, with two limiting values (by angular resolution) of
the multipoles $\ell_{max}\le150$ ($\theta\ge$36\arcmin) and
$\ell_{max}\le20$ ($\theta\ge$260\arcmin) according to the
decomposition into spherical \linebreak harmonics:
$$
\Delta T(\theta,\phi)= \sum_{\ell=2}^{\infty}\sum_{m=-\ell}^{m=\ell}
      a_{\ell m} Y_{\ell m} (\theta, \phi)\,,
$$
where the spherical harmonics is
$$
Y_{\ell m}(\theta,\phi) = \sqrt{{(2\ell+1)\over 4\pi}{(\ell-m)!
\over (\ell+m)!}}P_\ell^m(x) e^{i m\phi},
$$

\noindent $x=\cos\theta$\, $P_\ell^m(x)$ are the associated
Legendre polynomials, $\ell$ and $m$ are the multipole number and
its modes, respectively. For the expansion of spherical harmonics,
we used the GLESP package\footnote{\tt http://www.glesp.nbi.dk}
\cite{glesp:verkh2}.

\subsection{Signal Histograms in CMB Pixels}

\begin{figure*}[!th]
\centerline{\hbox{
\psfig{figure=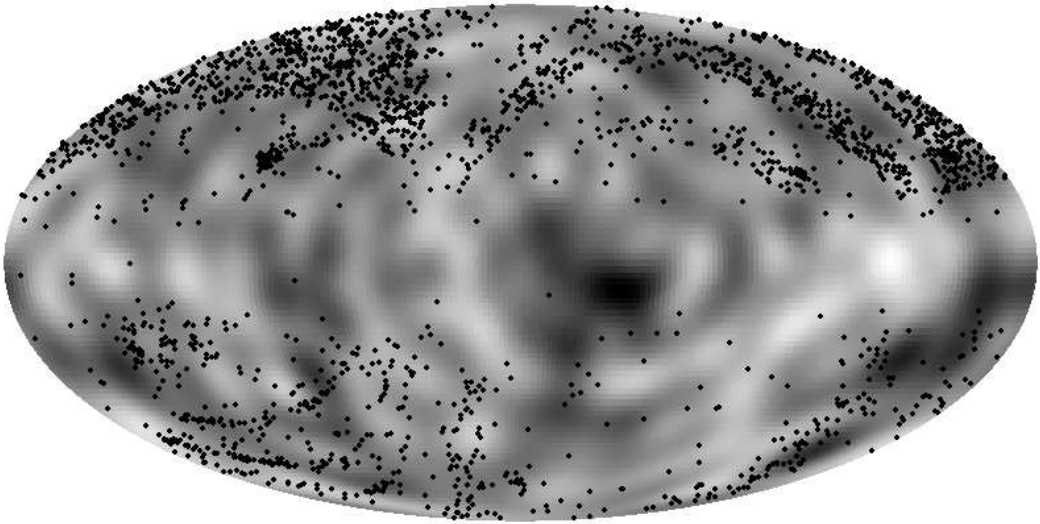,width=14cm}
}} \caption{Position of selected galaxies (black circles) in the
CMB WMAP map with a resolution of $\ell_{max}=20$ in galactic
coordinates. Dark spots in the map correspond to the cold signal,
while light spots---to the hot signal.} \label{f2:verkh2}
\end{figure*}

To analyze the Gaussianity of the distribution in the first order
and compare it with the model, we took the amplitude of CMB signal
fluctuations in the map pixels, onto which the radio galaxies are
projected (Fig.\,2), and constructed the histograms of its
distribution (Fig.\,3). One of the rules of galaxy selection in
our sample was the presence of a spectroscopic redshift, measured,
as a rule, in objects outside the Galactic plane, hence the lack
of sources in the central band	of Fig.\,2 (the Galactic plane) is
a selection effect. The pixel size of maps, used for the analysis,
at resolutions of  $\ell_{max}\le150$ and $\ell_{max}\le20$
amounts to 36\arcmin$\times$36\arcmin\, and
260\arcmin$\times$260\arcmin, respectively. The analysis was
carried out for maps with $\ell_{max}\le150$, since the ILC data
in the WMAP7 release are presented with a better resolution.

Figure\,3 shows the corresponding histograms for the CMB
fluctuation amplitude distributions of the ILC map in pixels,
corresponding to the direction to the catalog radio galaxies. The
histograms present a 1$\sigma$-spread, calculated by two methods
on the pixel statistics: (1) for the given 100 realizations of the
microwave background maps in the $\Lambda$CDM cosmological model
with homogeneous and isotropic Gaussian random fields, resulting
in the corresponding CMB fluctuations, and (2) for 100
realizations of a random arrangement of 2442 points in the ILC
WMAP7 map itself. The value of the histogram bean is 0.02\,mK.

\begin{figure*}[!th]
\centerline{ \vbox{
\hbox{
\psfig{figure=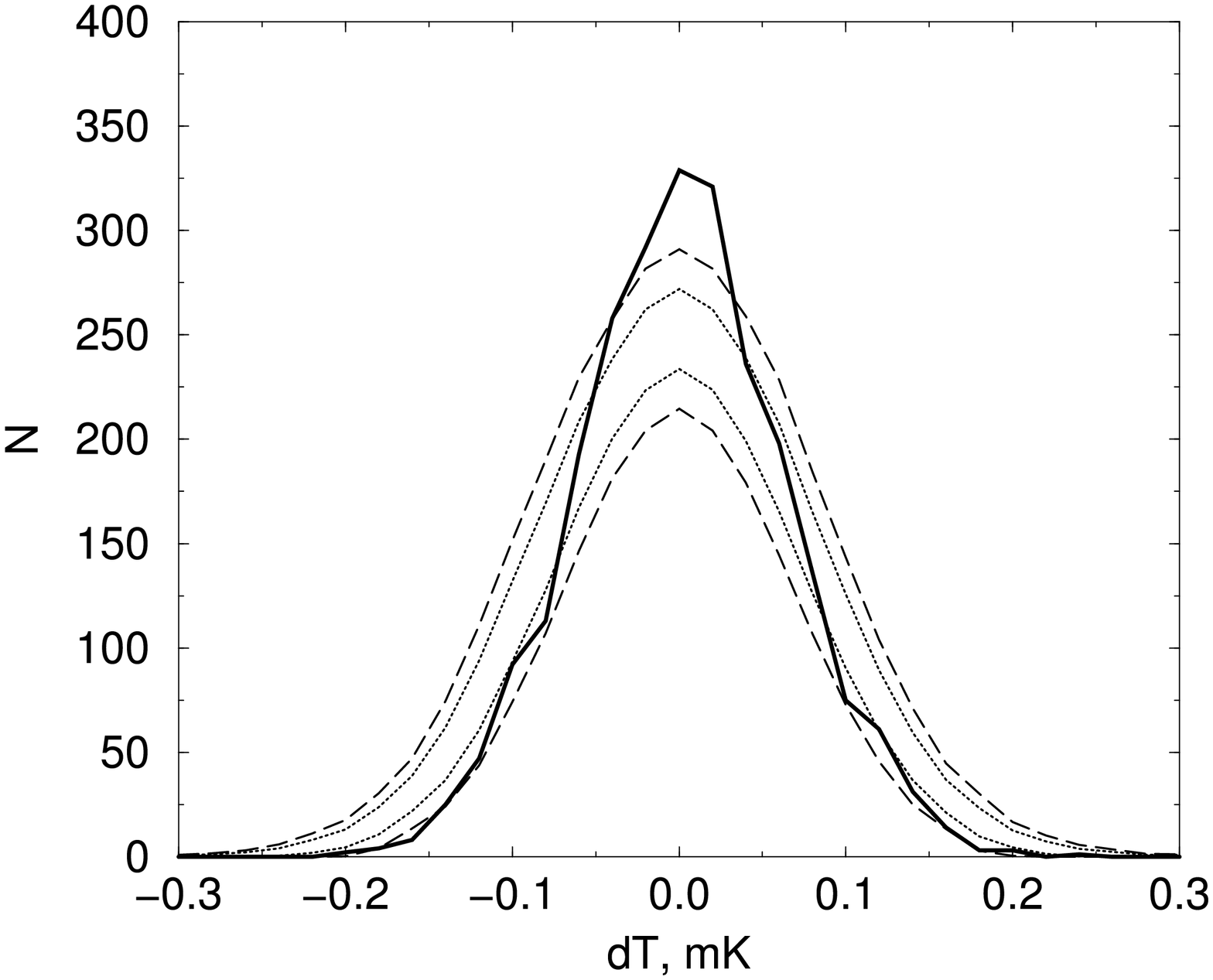,angle=-90,width=7cm}
\psfig{figure=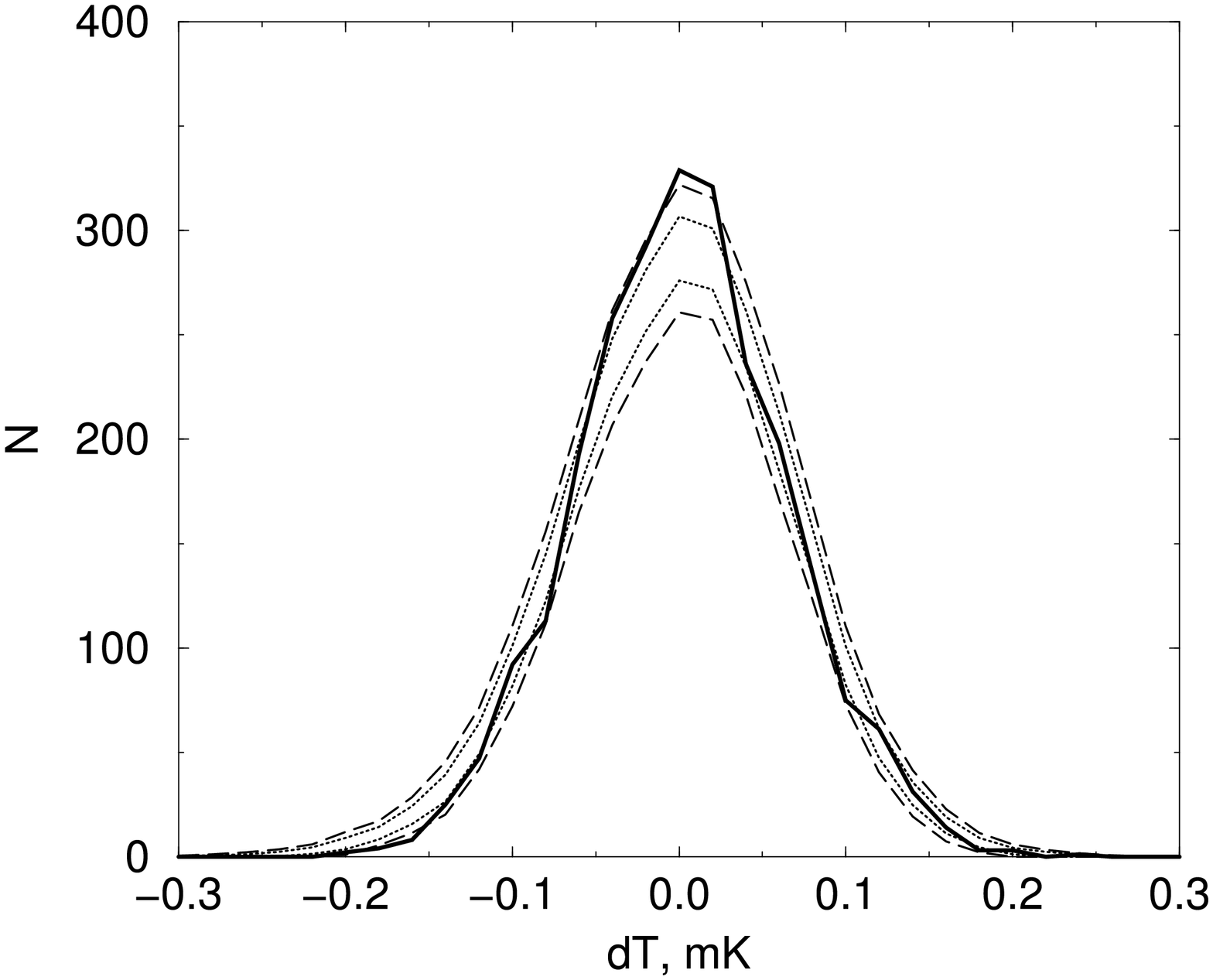,angle=-90,width=7cm}
} \hbox{
\psfig{figure=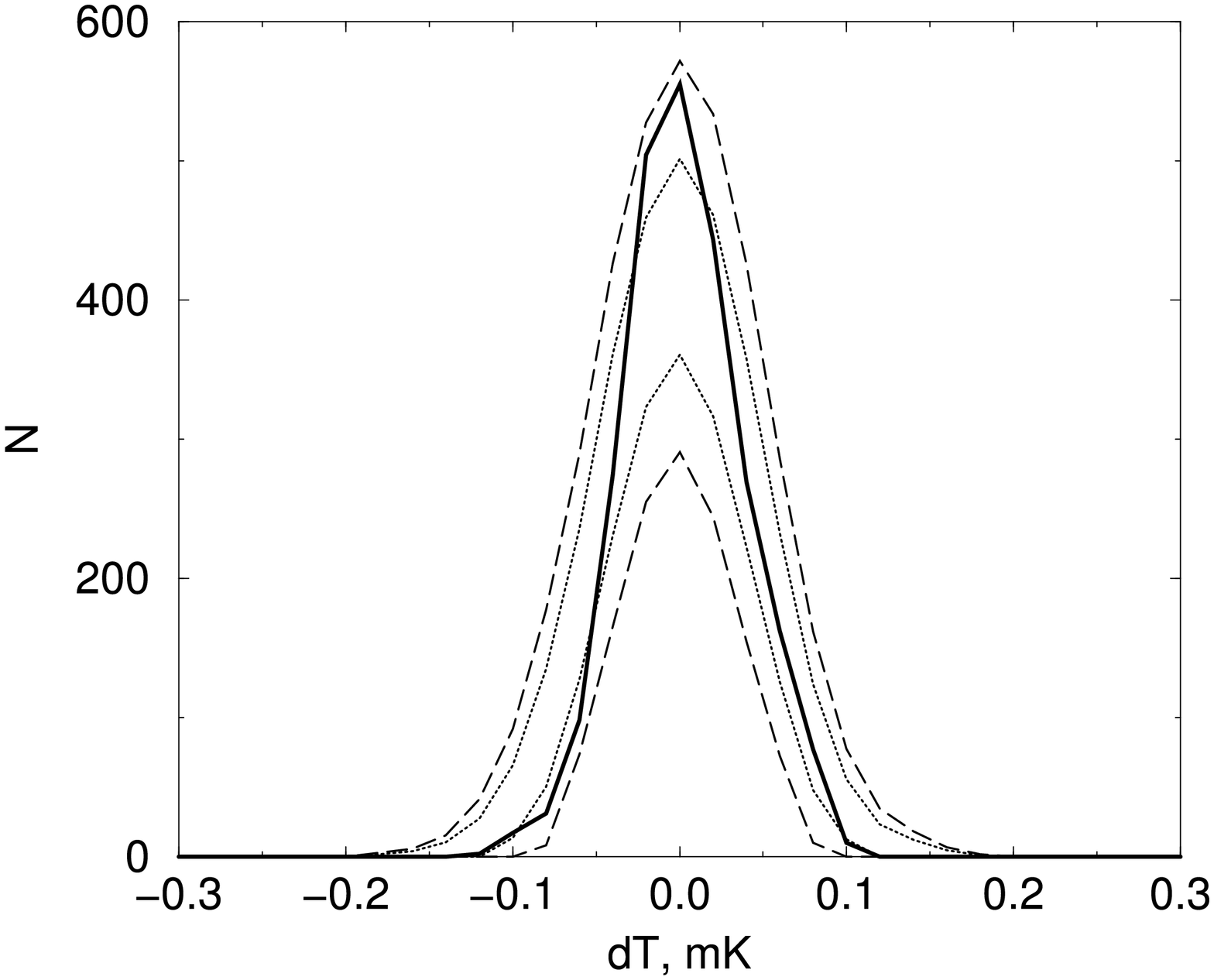,angle=-90,width=7cm}
\psfig{figure=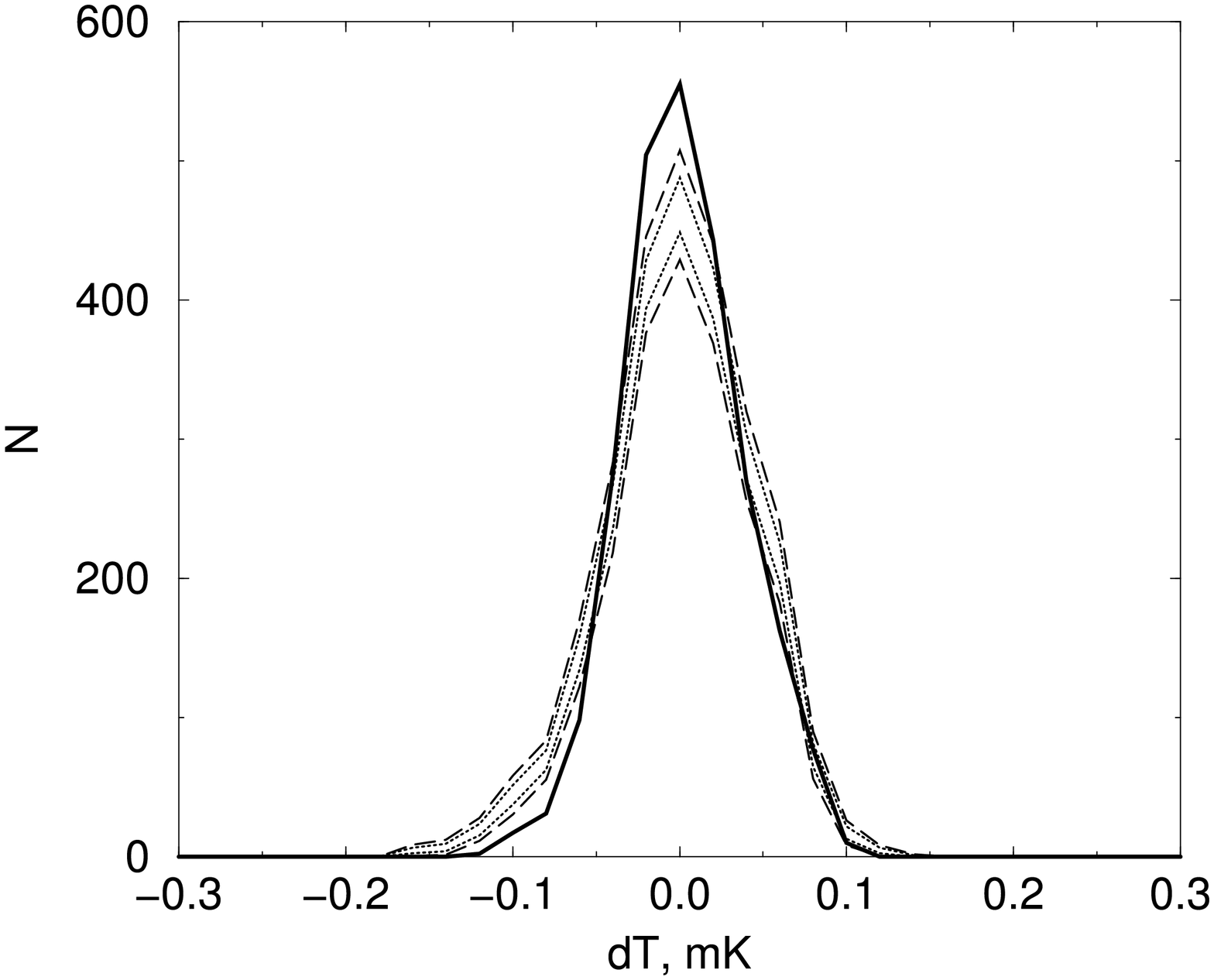,angle=-90,width=7cm}
} }} \caption{The histograms of the  distribution of the microwave
background signal value (the thick solid line) in the ILC map,
measured in pixels, corresponding to the direction to radio
galaxies. Above: histograms for the pixel map resolution of
$\ell_{max}\le150$. Bottom: histograms for the pixel map
resolution of $\ell_{max}\le20$. Left: the dotted and dashed lines
show the levels of $1\sigma$ and $2\sigma$-spreads, respectively,
calculated from the data of 100 realizations of the $\Lambda$CDM
cosmological model. Right: The dotted and dashed lines	show the
levels of $1\sigma$ and $2\sigma$-spreads, respectively,
calculated from the data of 100 realizations of a random
distribution of points in the ILC WMAP seven-year map.}
\label{f3:verkh2}
\end{figure*}

Note the particularities of the distributions shown in Fig.\,3.
All the signal variation distributions in the studied pixels are
close to normal. The distribution maximum $N$ is located strictly
in the region of zero amplitude fluctuations. Its value exceeds
the mean expected level by more than 3$\sigma$ for the maps with a
resolution of $\ell_{max}\le150$ and about $2\sigma$ with
$\ell_{max}\le20$ for the spread estimates, calculated using the
Monte Carlo method for Gaussian background fluctuations in the
$\Lambda$CDM cosmology (left graphs of the figure), and over
$2\sigma$ for the estimates, obtained in randomly selected pixels
of the ILC map.
The estimates are presented in the Table.

\begin{table}
\caption{{\bf Table.} Parameters of normal
distribution (the amplitude $N$ and the width parameter
\mbox{$s_0=\theta_{0.5}(8\ln 2)^{-1/2}$}), corresponding to the
distribution of the CMB temperature fluctuation amplitude in the
pixels of galaxy positions for maps with resolutions $\ell=150$
and $\ell=20$. The estimates of the distribution of temperature
minima and maxima at the $\sigma$-scatter are obtained when
modeling the background fluctuations in 100 realizations of a
random signal in the   $\Lambda$CDM cosmological model (marked as
$lcdm$), and in the simulations of 100 realizations of a random
scatter of galaxy locations in the ILC map (noted as $ilc$).}
\begin{tabular}{l|c|c}
\hline
Value		   & Amplitude & $s_0$, \\
Distribution		       &   $N$	  &   mK	\\
\hline
$T(\ell=150)$		   & 319.2  &  0.060	 \\
$T_{min}(\ell=150)^{lcdm}$ & 232.7  &  0.072	 \\
$T_{max}(\ell=150)^{lcdm}$ & 271.0  &  0.082	 \\
$T_{min}(\ell=150)^{ilc}$  & 276.1  &  0.064	 \\
$T_{max}(\ell=150)^{ilc}$  & 304.4  &  0.068	 \\
\hline
$T(\ell=20)$		   & 552.6  &  0.035	 \\
$T_{min}(\ell=20)^{lcdm}$  & 362.4  &  0.041	 \\
$T_{max}(\ell=20)^{lcdm}$  & 501.1  &  0.049	 \\
$T_{min}(\ell=20)^{ilc}$   & 438.8  &  0.041	 \\
$T_{max}(\ell=20)^{ilc}$   & 474.5  &  0.043	 \\
\hline
\end{tabular}
\end{table}

In addition, for both resolutions  $\ell=150$, and \linebreak
$\ell=20$, the ratio $(T_{max}/T_{min})^{lcdm}
> (T_{max}/T_{min})^{ilc}$, indicating a lower dispersion of signal variations
in the map than that, expected in the $\Lambda$CDM model.

\subsection{Dipole Estimation}

Using the value of temperature fluctuations in pixels in the map
with a resolution of $\ell=150$, applying the least-squares method
we estimated the corresponding dipole in the form

\small
$$
T(l,b) = T_x \cos(l)\cos(b) + T_y \sin (l) \cos(b) + T_z \sin(b),
$$
\normalsize

\noindent where $(l,b)$ are the galactic coordinates: longitude
and latitude, $T(l,b)$ is the value of CMB temperature
fluctuations in mJy, taken in the pixels corresponding to the
positions of radio galaxies, and  $(T_x,T_y,T_z)$ are the dipole
components, which are found to be equal to (0.0116, 0.0036,
0.0026), respectively.


Hence, we obtain the following for the extrema:
$$
T_{extrem} = \pm\sqrt{T_x^2+T_y^2+T_z^2},
$$
$$
b=\pm\arctan(T_z/\sqrt{T_x^2+T_y^2}),
$$
$$
l=\arctan(T_y/T_x), $$
\noindent or in the maximum
$T_{max}=0.0124$\,mK with coordinates $(l_{max}, b_{max})=$
$(17^\circ.307, 11^\circ.844)$, and in the minimum
$T_{min}=-0.0124$\,mK with coordinates \linebreak $(l_{min},
b_{min})=(197^\circ .307$, $-11^\circ .844)$. The dipole's
position on the sphere is shown in Fig.\,4.

\begin{figure*}[!th]
\centerline{\hbox{
\psfig{figure=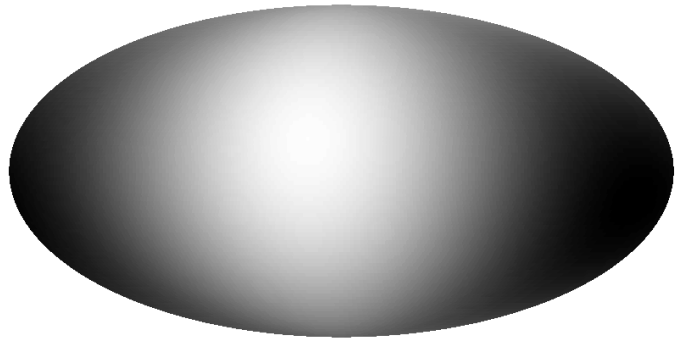,width=9cm}
}} \caption{The position of the dipole on the sphere, determined
with the least-squares method from the CMB pixel values in the
locations of distant radio galaxies, in galactic coordinates.}
\label{f4:verkh2}
\end{figure*}

Despite the fact that the estimated dipole is visually ``drawn''
towards the Galactic center, the significance of its position, as
well as the reality of its existence seems doubtful in view of its
small amplitude: $T_{max}=0.012$\,mK. This value lays within the
$\sigma$-scatter from the mean, i.e. within the noise, estimated
by  modelling of 50 CMB signal realizations in the $\Lambda$CDM
model, and determining the parameters of the dipole in each of the
realizations for the coordinates of the radio galaxy catalog. The
modelling results yield an average estimate of the parameters
$(T_x,T_y,T_z)=(0.003\pm 0.016, 0.001 \pm 0.018, -0.002\pm
0.010)$.

\section{RESULTS}

We investigated the properties of the  CMB \linebreak WMAP7 signal
in the fields of distant ($z>0.3$) radio galaxies. In general, the
signal distribution over the areas corresponds to normal.
Moreover, in this distribution the histogram amplitude is higher
than expected (more than by a $\sigma$-scatter) both for the
$\Lambda$CDM cosmological model and for the random positioning of
objects on the ILC WMAP map. That is, the pixels, corresponding to
the direction to distant radio galaxies, are dominated by small
(approaching zero) fluctuation values, which leads to an increase
in the histogram amplitude above the expected value. Absence of a
shift in the position of the histogram peak indicates either the
absence of any signal in these pixels, or a compensation of the SZ
effect by the radio emission of the galaxy (see, e.g.,
\cite{diego_partridge:verkh2}). Theoretically, one could expect a
shift either towards the negative signal in the case of the SZ
effect, or towards the positive signal in the case of existence of
a residuary signal from the radio source after the component
separation. In the presence of both factors we would see an
increase in the width of the histogram, compared with the simple
models of the background perturbation. However, an inverse effect
is observed: the resulting  CMB WMAP7 signal distribution,
measured in the regions of radio galaxy positions on the sphere is
in fact narrower than the one, expected for random Gaussian
fluctuations, and than the one, observed in the ILC map on the
average. In addition, the variance of the $s_0$ distribution is
smaller than expected in models, which also indicates an increased
number of pixels with zero signal in the fields of radio galaxies.
And if the correlation $(T_{max}/T_{min})^{lcdm} >
(T_{max}/T_{min})^{ilc}$ can be explained by an underestimated
value of the amplitude of the ILC quadrupole, then the cause of
the effect of signal attenuation in the fields of radio galaxies
is as yet unclear.

We as well tried to test the effect of existence of a dipole in
the CMB data  in the regions of galaxy clusters, discovered
~\mbox{in \cite{kash1:verkh2,kash2:verkh2}}. It is determined, as
the authors suppose, by the kinematic SZ effect, and associated
with the ``dark bulk flow'' of matter. To verify this phenomenon,
we constructed a dipole using the values in the pixels of galaxy
positions, and found that this dipole's amplitude is below the
$\sigma$ noise level of model variations.

Note that the size of the ILC map pixel in general does not allow
to valuably investigate the SZ effect. In addition, it is clear
that the nonzero dipole can almost always be defined for a finite
sample of pixels with nonzero values, what the models are
demonstrating.	Although we do not rule out the possibility that
for the nearby clusters of galaxies the background fluctuations
can show the existence of a common movement, distinct from the
known CMB kinematic dipole, but for distant objects, as in our
case, the Hubble flow will be dominating. We shall be able to
check this effect after the publication of maps of the Planck
mission.
\noindent
{\small

\section{ACKNOWLEDGMENTS}
The authors are grateful to P.~D.~Naselsky for helpful
discussions, to S.~A.~Trushkin for valuable comments that allowed
to improve the text and to O.~Nasonova for her aid with the
calculations in the MATLAB package. In the study we used the NED
database of extragalactic objects
The authors also used the CATS \cite{cats3:verkh2} database of
radio astronomy catalogs, the FADPS\footnote{\tt
http://sed.sao.ru/$\sim$vo/fadps\_e.html}
\cite{fadps:verkh2,fadps2:verkh2} system for processing the radio
astronomy data,	 and the GLESP package	data analysis of microwave
radiation on the \mbox{sphere \cite{glesp1:verkh2,glesp2:verkh2}.}
This work was supported by the Leading Scientific Schools of
Russia (S.\,M.\,Khaikin school) grant, and the RFBR grants
(project nos. \linebreak 09-02-00298,  08-02-00486). O.V.V. is
also grateful for the partial support of the Dmitry Zimin Dynasty
Foundation.

\end{document}